\documentclass[preprint]{aastex}
\usepackage{epsf,graphics}
\shorttitle{Neutrino-cooled accretion disk in GRBs}
\shortauthors{Kohri \& Mineshige}

\newcommand{\mev}{ {\rm MeV} }

\newcommand{\mb}{ {\rm mb} }
\newcommand{\g}{ {\rm g} }
\newcommand{\cm}{ {\rm cm} }
\newcommand{\s}{ {\rm s} }
\newcommand{\K}{ {\rm K} }
\newcommand{\kb}{ k_{\rm B} }
\newcommand{\erg}{ {\rm erg} }
\newcommand{\mbh}{ {M_{\rm BH}} }
\newcommand{\order}{{\cal O}}
\newcommand{\anti}[1]{{\overline{#1}}}

\newcommand{\gsim}{\mbox{\raisebox{-1.0ex}{$\stackrel{\textstyle >}
{\textstyle \sim}$ }}} 
\newcommand{\lsim}{\mbox{\raisebox{-1.0ex}{$\stackrel{\textstyle <}
{\textstyle \sim}$ }}}  
\newcommand{\gtsima}{$\; \buildrel > \over \sim \;$}
\newcommand{\ltsima}{$\; \buildrel < \over \sim \;$}
\newcommand{\simgt}{\lower.5ex\hbox{\gtsima}}
\newcommand{\simlt}{\lower.5ex\hbox{\ltsima}}

\begin{document}




\title{Can neutrino-cooled accretion disk be an origin of gamma-ray
bursts?}

\author{Kazunori Kohri and Shin Mineshige}
\affil{
Yukawa Institute for Theoretical Physics, Kyoto University, 
Kyoto, 606-8502, Japan
}
\email{kohri@yukawa.kyoto-u.ac.jp, minesige@yukawa.kyoto-u.ac.jp}

\received{2001}
\accepted{2001}

\begin{abstract}
    It is often considered that a massive torus with solar mass or so
    surrounding a stellar-mass black hole may be a central engine of a
    gamma-ray burst.  We study the properties of such massive
    accretion tori (or disks) based on the $\alpha$ viscosity model.
    For surface density exceeding about $10^{20}$ g~cm$^{-2}$, which
    realizes when about a solar-mass material is contained within a
    disk with a size of $\sim 5 \times 10^6$ cm, we find that (1)
    luminosity of photons is practically zero due to significant
    photon trapping, (2) neutrino cooling dominates over advective
    cooling, (3) pressure of degenerate electrons dominates over
    pressure of gas and photons, and (4) magnetic field strength
    exceeds the critical value of about $4 \times 10^{13}$ G, even if
    we take 0.1 \% of the equi-partition value.  The possible observable
    quantum electrodynamical (QED) effects arising from super-critical
    fields are discussed.  Most interestingly, photon splitting may
    occur, producing significant number of photons of energy below
    $\sim 511$ keV, thereby possibly suppressing e$^\pm$ pair
    creation.
\end{abstract}

\setcounter{equation}{0}
\section{Introduction}
\label{sec:intro}

Research on gamma-ray bursts (GRBs) has made rapid progress
over the past decade, particularly after the discovery of afterglow
by Beppo-SAX (see recent reviews by 
M\'esz\'aros, Rees, \& Wijers 1999; Piran 2000; M\'esz\'aros 2001).
The current common belief is that GRBs are the most energetic explosions
which ever occurred in the Universe, releasing
energy of $\sim 10^{51}-10^{53}$ erg only in a few to few tens of seconds.
It is also widely argued that GRBs result from the
conversion of the kinetic energy of ultra-relativistic particles
created within a fire ball.
A relativistic fireball shock model 
was proposed to resolve the compactness problem and
has had some success in explaining 
how the GRB and afterglow radiation arise
 (Rees \& M\'esz\'aros 1992, 1994; Wijers \& Galama 1999).

However, the central engines of GRBs creating initial hot plasma
or extremely energetic particles
are not well understood yet due mainly to the fact that
they are hidden from our view.
It is usually argued that relativistic phenomena should be somehow 
involved, since otherwise it is difficult to explain
huge fluence and rapid burst-like bust profiles.
(Narayan, Paczy\'nski \& Piran 1992).  Along this line,
many interesting possibilities have been proposed so far, 
such as
(1) mergers of double neutron star (NS) binaries or NS and BH (black hole)
    (Paczy\'nski 1986, 1991; Goodman 1986; Eichler et al. 1989), 
(2) BH-WD (white dwarf) merger (Fryer et al. 1999);
(3) BH-helium star merger (Fryer et al. 1999);
(4) failed supernova (or collapsars, Woosley 1993; Paczy\'nski 1998;
    MacFadyen \& Woosley 1999), 
and (5) magnetar, rapidly spinning neutron star 
with extremely large magnetic fields (Usov 1992, 1994; 
Thompson 1994; Spruit 1999).
It is of great importance to note that
almost all the models (except for the magnetar hypothesis)
predicts a similar configuration as an end result; namely,
formation of a few solar mass black hole surrounded by
a temporary debris torus (or disk) with mass of $0.01 - 1.0~M_\odot$
whose accretion can provide a sudden
release of gravitational energy (M\'esz\'aros, Rees, \& Wijers 1999).
Then, the key issue will be to understand the properties 
of such a compact and massive disk 
associated with a huge mass accretion rate.

Accretion models for gamma-ray bursts were first considered by
Narayan, Paczy\'nski \& Piran (1992), and recently discussed in more
details by Popham, Woosley, \& Fryer (1999) and Narayan, Piran, \&
Kumar, (2001, hereafter NPK).  According to them, neutrino cooling
dominates over radiative loss at very huge accretion rates (see,
however, Chevalier 1996; see also Ruffert \& Janka 1999 for a
numerical simulation of a binary merger including neutrino losses.
Therefore, such disks (or flows) are called as neutrino-dominated
accretion flow (NDAF).  In the present study, we elucidate the theory
of NDAF based on the $\alpha$ viscosity prescription, paying special
attention to their thermal structure and the implications of expected
huge magnetic fields.  We present the resultant equilibrium solutions
and properties of neutrino-cooled disks in section 2.  We then discuss
the effects of huge magnetic fields and implications on neutrino
emission in section 3.  Final section is devoted to conclusions.

\section{Structure of Neutrino-Cooled Disk}
\label{sec:model}
\subsection{$\Sigma-T$ diagram}

In this study we adopt the following basic equations
based on the Newtonian dynamics, for simplicity
(see, e.g., Kato, Fukue, \& Mineshige 1998). 
We use the cylindrical coordinates, $(r,\varphi,z)$.  Then,
the mass conservation in the vertical and radial direction is,
respectively, represented by
\begin{eqnarray}
    \label{eq:continuity}
    \rho = \frac{\Sigma}{2 H} \qquad{\rm and}\qquad
     \dot{M} = - 2\pi r v_r \Sigma,
\end{eqnarray}
where $\rho$ is the matter density,
$\Sigma$ is the surface density of the disk, $H$ is the disk
half-thickness, $\dot{M}$ is the mass accretion rate, and $v_r$ is the
radial velocity (negative for inflow). The disk half-thickness is,
from hydrostatic balance in the vertical direction, given by
\begin{eqnarray}
    \label{eq:disk-half-thickness}
    H = c_{\rm s} / \Omega,
\end{eqnarray}
where the (isothermal) sound velocity $c_{\rm s}$ is defined by
\begin{eqnarray}
    \label{eq:sound-velocity}
    c_{\rm s} \equiv \sqrt{p / \rho},
\end{eqnarray}
with the pressure $p$.  The angular velocity, $\Omega$, is
in the Newtonian approximation given by
\begin{eqnarray}
    \label{eq:angular_velocity}
    \Omega = \sqrt{G M_{\rm BH}/ r^3},
\end{eqnarray}
with $G$ being the gravitational constant 
and $M_{\rm BH}$ being the black-hole mass.  
The pressure is composed of three terms:
\begin{eqnarray}
    \label{eq:total_pressure}
    p = p_{\rm rad} + p_{\rm gas} + p_{\rm d},
\end{eqnarray}
where $p_{\rm rad}$ is the radiation pressure, $p_{\rm gas}$ is the
gas pressure of the nonrelativistic and non-degenerate particles, and
$p_{\rm d}$ is the pressure of the degenerate particles.  The
radiation pressure is
\begin{eqnarray}
    \label{eq:rad_pressure}
    p_{\rm rad}=\frac{a}{6}g_* T^4,
\end{eqnarray}
where $g_*$ is statistical degree of freedom in the radiation (=2 for
$\gamma$, or 11/2 for $\gamma$ and $e^+e^-$ plasma) and
$a$ is the radiation constant.
The gas pressure is
\begin{eqnarray}
    \label{eq:gas_pressure}
    p_{\rm gas} = \sum_i n_i k_{\rm B} T,
\end{eqnarray}
where $k_{\rm B}$ is the Boltzmann constant, $n_i$ is the number
density of particle ``$i$'', and the suffix $i$ runs the
nonrelativistic and non-degenerate particles at the temperature (e.g.,
$e^-$ and nucleons at $T \lsim m_{\rm e} c^2/k_{\rm B}$). The pressure
of the degenerate particles is written as
\begin{eqnarray}
    \label{eq:tot_degeneracy_pressure}
    p_{\rm d} = p_{\rm d, nonrel} + p_{\rm d, rel}.
\end{eqnarray}
Then, 
\begin{eqnarray}
    \label{eq:nonrel_degeneracy_pressure}
    p_{\rm d, nonrel} = \sum_{i={\rm nonrel}}
    \frac15\left(\frac{6\pi^2}{g_i}\right)^{2/3}\frac{\hbar^2n_i^{5/3}}{m_i},
\end{eqnarray}
is for nonrelativistic particles which satisfy the condition of the
degeneracy,
\begin{eqnarray}
    \label{eq:deg_con_nonrela}
    n_i \gg g_i(m_i k_{\rm B} T/2 \pi \hbar^2)^{3/2},
\end{eqnarray}
for $T \ll m_i c^2/k_{\rm B}$ and $\mu_i \ll m_i c^2$ with their
mass $m_i$, statistical degree of freedom $g_i$ and their chemical
potential $\mu_i$, and
\begin{eqnarray}
    \label{eq:rel_degeneracy_pressure}
    p_{\rm d, rel} = \sum_{i={\rm rel}}\frac14
    \left(\frac{6\pi^2}{g_i}\right)^{1/3}\hbar^2c n_i^{4/3},
\end{eqnarray}
is for relativistic particles which satisfy the condition of the
degeneracy, 
\begin{eqnarray}
    \label{eq:deg_con_rela}
    n_i \gg g_i(k_{\rm B} T/\hbar c)^{3}/\pi^2,
\end{eqnarray}
for $T \gg m_i c^2/k_{\rm B}$ or $\mu_i \gg m_i c^2$~.
In this study, for simplicity we assume the complete degeneracy
whenever the above condition of the degeneracy is satisfied.

In Fig.~\ref{fig:rho_cont}, we plot the contours of the matter density
on the ($\Sigma$, $T$) plane. From the figure we can read off the
matter density.  It also discriminates the regions where each pressure
component is dominant and also the regions where electrons and/or
nucleons are degenerate.  We understand that degeneracy pressure is
important at high density and low temperature regimes.  Such regimes
inevitably appear in very massive disks, and one can never neglect the
contributions by degeneracy pressure in GRBs unlike the cases of
binary systems or galactic centers.  We also notice that one should
distinguish the regime where only electrons are degenerate from the
one where both electrons and nucleons are degenerate.

\subsection{chemical potential of electrons}

When electrons are degenerate, there emerges an important consequence
on the process of neutrino emission, that is, e.g., since the pair
creation of electron and positron in the electromagnetic thermal bath
is suppressed by the charge neutrality (see below), the neutrino
emission is also suppressed (This is not discussed in NPK). The
electron chemical potential $\mu_e$ is determined by the condition of
charge neutrality,
\begin{eqnarray}
    \label{eq:mu_e}
    n_p = n_{e^-} - n_{e^+},
\end{eqnarray}
with the number density of electron and positron,
\begin{eqnarray}
    \label{eq:n_e}
    n_{e^-} &=& \frac{1}{\hbar^3\pi^2}\int^{\infty}_{0}dp p^2
    \frac{1}{e^{(\sqrt{p^2c^2 + m_e^2c^4} - \mu_e)/k_{\rm B}T} + 1}, \\
    n_{e^+} &=& \frac{1}{\hbar^3\pi^2}\int^{\infty}_{0}dp p^2
    \frac{1}{e^{(\sqrt{p^2c^2 + m_e^2c^4} + \mu_e)/k_{\rm B}T} + 1}.
\end{eqnarray}
In the following sections, it is convenient to introduce a
dimensionless chemical potential of electrons which is defined by
\begin{eqnarray}
    \label{eq:eta_e}
    \eta_e = \mu_e / k_{\rm B}T.
\end{eqnarray}
Then, the complete degeneracy of electrons corresponds to $\eta_e \gg
1$, and the charge neutrality in Eq.~(\ref{eq:mu_e}) becomes
\begin{eqnarray}
    \label{eq:charge_complete_edeg}
    n_p \simeq n_{e^-} \simeq \frac1{3\pi^2}\left(\frac{\mu_e}{\hbar
      c}\right)^3.
\end{eqnarray}

The chemical potentials of protons and neutrons are defined through
the relation,
\begin{eqnarray}
    \label{eq:n_N_def}
    n_{p} &=& \frac{1}{\hbar^3\pi^2}\int^{\infty}_{0}dp p^2
    \frac{1}{e^{(\sqrt{p^2c^2 + m_p^2c^4} - \mu_p)/k_{\rm B}T} + 1}, \\
    n_{n} &=& \frac{1}{\hbar^3\pi^2}\int^{\infty}_{0}dp p^2
    \frac{1}{e^{(\sqrt{p^2c^2 + m_n^2c^4} - \mu_n)/k_{\rm B}T} + 1}.
\end{eqnarray}
If the nucleons are not degenerate, we have
\begin{eqnarray}
    \label{eq:n_N_non_deg}
    n_{i} &=& 2 \left(\frac{m_i\kb
      T}{2\pi\hbar^2}\right)^{3/2}e^{-(m_ic^2 -
    \mu_i)/\kb T}, 
\end{eqnarray}
for $i = p$, and $n$.  On the other hand, if nucleons are degenerate,
we have
\begin{eqnarray}
    \label{eq:eq:n_N_deg}
    n_i = \frac1{3\pi^2}
    \left(\frac{2\tilde{\mu}_im_i}{\hbar^2}\right)^{3/2},
\end{eqnarray}
for the complete degenerate limit with
\begin{eqnarray}
    \label{eq:tilde_chemi}
    \tilde{\mu}_i = \mu_i - m_ic^{2}.
\end{eqnarray}

In this study, we did not directly solve the exact equation in
Eq.~(\ref{eq:mu_e}) to derive the chemical potential of
electrons. Instead, we performed the following approximated
computations in the complete-degeneracy regime ($\eta_e \gg 1$). As
will be shown later, nucleons are in $\beta$-equilibrium at a high
temperature $T \gtrsim 5.0 \times 10^{10} \K
(\alpha/0.1)^{1/5}(r/4r_g)^{-3/10}(\mbh/3M_{\odot})^{-1/5}$ where the
neutrino emissions are effective and mainly contribute to the cooling
process. Then, the chemical equilibrium is realized,
\begin{eqnarray}
    \label{eq:chem_equiv}
    \mu_p + \mu_e = \mu_n,
\end{eqnarray}
or 
\begin{eqnarray}
    \label{eq:chem_equiv2}
    \tilde{\mu}_p + \mu_e - Q = \tilde{\mu}_n,
\end{eqnarray}
where $Q = (m_n - m_p)c^2 \simeq 1.29$ MeV. In the regime where only
electrons are degenerate, $\beta$-equilibrium means
\begin{eqnarray}
    \label{eq:beta_equiv_edeg}
    \frac{n_p}{n_n} = \exp{[(Q - \mu_e)/(k_{\rm B}T)]}.
\end{eqnarray}
The electron fraction
$Y_e \equiv n_p/(n_p + n_n) $ is related with the above ratio by $Y_e
= 1/(1+(n_p/n_n)^{-1})$. Combined with the condition of the charge
neutrality, i.e., Eq.~(\ref{eq:charge_complete_edeg}), we may only
solve the equation,
\begin{eqnarray}
    \label{eq:solve_chemi_edeg}
    \frac{\rho}{m_p} \exp(-\mu_e/(\kb T)) =
    \frac1{3\pi^2}\left(\frac{\mu_e}{\hbar c}\right)^3,
\end{eqnarray}
for $\mu_e \gg k_{\rm B}T \gg Q$. On the other hand, it is relatively
easy for us to calculate $\mu_e$ in the regime where both electrons
and nucleons are degenerate. From the charge neutrality
(Eq.~(\ref{eq:charge_complete_edeg})) and chemical equilibrium
(Eq.(\ref{eq:chem_equiv2})), we find that $\mu_e \sim \tilde{\mu}_n
\gg \tilde{\mu}_p$ for $\kb T \ll \mu_e < \order(10^2) \ \mev$. Then,
from Eq.~(\ref{eq:eq:n_N_deg}) we can approximately estimate the
chemical potential of electrons as
\begin{eqnarray}
    \label{eq:solve_chemi_Ndeg}
    \mu_e &\simeq& \tilde{\mu_n} \\
    &\simeq& 6.628 \ \mev \ (\rho/10^{13} \g~\cm^{-3})^{2/3},
\end{eqnarray}
in the complete-degeneracy limit of both electrons and nucleons.

We also plot the contours of the chemical potential of electrons in
units of MeV ($\mu_{e, {\rm MeV}}$ = 10, and 20) in
Fig.~\ref{fig:rho_cont}.  The chemical potential of electrons play
important roles to estimate the emission rates of neutrinos which will
be discussed in the next subsection.

\subsection{Heating and cooling rates}
\label{sec:heating-cooling-rates}

The most important relation in the accretion disk theory is the energy
balance between the heating and cooling processes.  Distinct branches
(such as a standard disk or a slim disk, etc.) appear because of
different heating and/or cooling sources being dominant.  According
to the standard $\alpha$ viscosity model, we express the vertically
integrated heating rate (over a half thickness, $H$) as
\begin{eqnarray}
    \label{eq:Qvis}
    Q^+ = Q^+_{\rm vis} 
        = \frac{9}{8}\nu\Sigma\Omega^2,
\end{eqnarray}
where $Q^+_{\rm vis}$ denotes the viscous heating rate per unit
surface area, the kinetic viscosity $\nu$ is related to the viscosity
parameter $\alpha$ by
\begin{eqnarray}
    \label{eq:kinematic_viscosity}
    \nu = \frac23\alpha c_{\rm s} H.
\end{eqnarray}
If we assume the angular momentum conservation, we obtain
\begin{eqnarray}
    \label{eq:angular_momentum}
     \nu \Sigma = \frac{\dot M}{3\pi}
        \left(1-\sqrt{\frac{r_{\rm in}}{r}}\right),
\end{eqnarray}
where $r_{\rm in}$ is the radius of the inner edge of the disk
(=~3~$r_{\rm g}$).  From Eqs.~(\ref{eq:Qvis}),~(\ref{eq:kinematic_viscosity})
and~(\ref{eq:angular_momentum}), we can relate the heating rate with
the mass accretion rate;
\begin{eqnarray}
    \label{eq:Q_mdot}
    \dot{M}= \frac{8\pi}{3}\Omega^{-2} 
                \left(1-\sqrt{\frac{r_{\rm in}}{r}}\right)^{-1}Q^+.
\end{eqnarray}

The cooling rate is, on the other hand,
summation of three major contributions;
\begin{eqnarray}
    \label{eq:cooling}
    Q^- = Q^-_{\rm rad} + Q^-_{\rm adv} + Q^-_{\nu},
\end{eqnarray}
where $Q^-_{\rm rad}$ is the radiative cooling rate, $Q^-_{\rm adv}$
is the advective energy transport (Abramowicz et al. 1988), and
$Q^-_{\nu}$ is the cooling rate due to neutrino loss. Note that
instead of including advective energy transport in the energy equation
NPK adopted the view of CDAF (convection-dominated accretion flow;
Igumenshchev, Abramowicz, \& Narayan 2000, Narayan, Igumenshchev, \&
Abramowicz, 2000, Quataert \& Gruzinov 2000).  We, in the present
study, retain the classical picture based on the (vertically) one-zone
treatment (e.g., Kato, Fukue, \& Mineshige 1998), since it is not yet
clear if CDAF provides precise description to the flow structure not
only in the low-luminosity regimes but also in the hyper-accretion
regimes.  However, such a distinction is not essential here, since we
are concerned with the regimes of even higher accretion rates, in
which neutrino emission is substantial (discussed later).

The radiative cooling rate is
\begin{eqnarray}
    \label{eq:cooling2}
    Q^-_{\rm rad} = \frac{g_* \sigma_{\rm s} T^4}{2 \tau_{\rm tot}},
\end{eqnarray}
where $\sigma_{\rm s}=\pi^2 k_{\rm B}^4/(60 \hbar^3 c^2)$ 
is the Stefan-Boltzmann constant and 
the optical depth, $\tau_{\rm tot}$, is given by
\begin{eqnarray}
    \label{eq:optical_depth}
    \tau_{\rm tot} = \kappa_{\rm R}\rho H = \frac{\kappa_{\rm R}\Sigma}{2},
\end{eqnarray}
with $\kappa_{\rm R}$ being the Rosseland-mean opacity,
\begin{eqnarray}
    \label{eq:opacity}
    \kappa_{\rm R} = 0.40 + 0.64 \times 10^{23}
    \left(\frac{\rho}{\g~\cm^{-3}}\right)\left(\frac{T}{\K}\right)^{-3}
    \g^{-1}~ \cm^2. 
\end{eqnarray}
The advective cooling rate is given by (Kato, Fukue, \& Mineshige
1998),
\begin{eqnarray}
    \label{eq:advective_cooling}
    Q^-_{\rm adv}=\Sigma T v_r \frac{d s}{dr},
\end{eqnarray}
where the radial velocity is $v_r = -\dot{M}/(2\pi r \Sigma)$
and $s$ denotes entropy per particle,
\begin{eqnarray}
    \label{eq:entropy_per_matter}
    s = \left(s_{\rm rad} + s_{\rm gas}\right)/\rho.
\end{eqnarray}
Here, the entropy density of the radiation is 
\begin{eqnarray}
    \label{eq:radiation-entropy}
    s_{\rm rad} = \frac{2}{3} a g_* T^3,
\end{eqnarray}
and entropy density of the gas (i.e., nonrelativistic particles) is
\begin{eqnarray}
    \label{eq:gas-entropy}
    s_{\rm gas} = \sum_i n_i\left(\frac52 +
      \ln\left[\frac{\g_i}{n_i}\left(\frac{m_i
          T}{2\pi}\right)^{3/2}\right]\right),
\end{eqnarray}
where the suffix $i$ runs over nonrelativistic nucleon and electron,
$g_i$ is the statistical degree of freedom of the particle.  In the
following, we approximated $ds/dr$ as $s/r$. Note that since the
entropy of degenerate particles is small in the complete-degeneracy
limit, we neglect it here.

The neutrino cooling rate is composed of four terms;
\begin{eqnarray}
    \label{eq:neutrino_cooling}
    Q^-_{\nu} = (\dot{q}_{Ne} + \dot{q}_{e^+e^-} + \dot{q}_{\rm brems}
    + \dot{q}_{\rm plasmon}) H,
\end{eqnarray}
where $\dot{q}_{Ne}$ is the electron-positron capture rate by a
nucleon ``N'', $\dot{q}_{e^+e^-}$ is the electron-positron pair
annihilation rate, $\dot{q}_{\rm brems}$ is the nucleon-nucleon
bremsstrahlung rate, and $\dot{q}_{\rm plasmon}$ is the rate of
plasmon decays. Note that NPK considered only the case of the
non-degenerate electrons in first two terms on the right-hand side of
Eq.~(\ref{eq:neutrino_cooling}). However, the case of the degenerate
electrons in the first two terms and the other two terms could be also
important.

The electron-positron capture rate is represented by two terms:
\begin{eqnarray}
    \label{eq:e-capture}
    \dot{q}_{Ne} = \dot{q}_{p+e^- \to n + \nu_e} 
    + \dot{q}_{n+e^+ \to p + \anti{\nu}_e},
\end{eqnarray}
with
\begin{eqnarray}
    \label{eq:e-p}
    \dot{q}_{p+e^- \to n + \nu_e}&=&\frac{G_{\rm F}^2}{2\pi^3
    \hbar^3c^2} (1 + 3 g_A) n_p \int_Q^{\infty} dE_e E_e \sqrt{E_e^2 -
    m_e^2c^4}
    \left(E_e - Q\right)^3 \frac1{e^{\left(E_e - \mu_e\right)/k_{\rm B}T}+1}, \\
    \label{eq:e+n}
    \dot{q}_{n+e^+ \to p + \anti{\nu}_e}&=&\frac{G_{\rm F}^2}{2\pi^3
    \hbar^3c^2} (1 + 3 g_A) n_n \int_{m_ec^2}^{\infty} dE_e E_e
    \sqrt{E_e^2 - m_e^2c^4} \left(E_e + Q\right)^3 \frac1{e^{\left(E_e
      + \mu_e\right)/k_{\rm B}T}+1},
\end{eqnarray}
where $G_{\rm F}$ is the Fermi coupling constant ($= 2.302 \times
10^{-22} \cm~\mev^{-1}$), and $g_A$ is the axial-vector coupling
constant of nucleon ($\sim$ 1.39) which is normalized by the
experimental value of neutron lifetime $\tau_n \simeq 886.7$ s. In the
non-degeneracy limit ($\mu_e \ll k_{\rm B}T$), it is easily estimated by
\begin{eqnarray}
    \label{eq:nucapture}
    \dot{q}_{Ne}=9.2\times 10^{33} \erg~ \cm^{-3}\s^{-1}
    \left(\frac{T}{10^{11}\K}\right)^6\left(\frac{\rho}{10^{10} \g~
      \cm^{-3}}\right).
\end{eqnarray}
On the other hand, it is a little complicated to estimate the electron
capture rate in the electron-degeneracy regime. In the only
electron-degeneracy regime, we have
\begin{eqnarray}
    \label{eq:only-e}
    \dot{q}_{Ne} = 1.1\times 10^{31} \eta_e^9 \ \erg~\cm^{-3}~\s^{-1}
    \left(\frac{T}{10^{11}\K}\right)^9,
\end{eqnarray}
for complete degeneracy limit of electrons. Note that in this limit,
independently we can also obtain the chemical potential of electrons
by the balance between the two dominant weak interaction rates, i.e.,
$\dot{q}_{p+e^- \to n + \nu_e} = 5.3\times 10^{30} \eta_e^9 \ 
\erg~\cm^{-3}~\s^{-1}(T/10^{11} \K)^9$ and $\dot{q}_{n+e^+ \to p +
\anti{\nu}_e} = 2.6 \times 10^{37} \exp(-m_e/(\kb T)) \exp(-\eta_e)
\erg~\cm^{-3}~\s^{-1} (\rho/10^{13} \g~\cm^{-3})(T/10^{11} \K)^6
$. Compared with the chemical potential which was obtained in
Eq~(\ref{eq:solve_chemi_edeg}) where we assumed the complete
$\beta$--equilibrium, they agree with each other just within a few
percent, e.g., $\eta_e \simeq 3.5$ at $\rho = 10^{13} \g~\cm^{-3}$ and
$T$ = 10 MeV/$\kb$. Therefore, our assumptions of the
complete-degeneracy limits would be reasonable.

When both electrons and nucleons are degenerate, the reactions $p+e^-
\leftrightarrow n + \nu_e$ are suppressed by the Fermi blocking of
degenerate nucleons in the final states. Then, for example the $dE_e$
integration in Eq.~(\ref{eq:e-p}) is limited as [$\mu_e$, $\infty$],
and we approximately obtain the relatively small cooling rate,
$\dot{q}_{Ne} = 5.0 \times 10^{32} \eta_e^7 \ 
\erg~\cm^{-3}~\s^{-1}({T}/{10^{11} \K})^9$.  Compared with
Eq.~(\ref{eq:only-e}), of course, we find that this rate is smaller by
a factor of $1/\eta_e^2$ in the complete degeneracy regime, i.e.,
$\eta_e \gg 1$.

The electron-positron pair annihilation rate through $e^+ + e^- \to
\nu + \anti{\nu}$ is
\begin{eqnarray}
    \label{eq:eeannihilation}
    \dot{q}_{e^+e^-}=4.8\times 10^{33} \erg~ \cm^{-3}\s^{-1}
    \left(\frac{T}{10^{11}\K}\right)^9,
\end{eqnarray}
in the non-degeneracy regime, e.g., see Itoh et al. (1989, 1990). When
the electrons are degenerate, the electron-positron pair annihilation
rate is too small to compete the other cooling process, and we can
neglect it.

The nucleon-nucleon bremsstrahlung rate through $ n + n \to n + n +
\nu + \anti{\nu}$ is represented by
\begin{eqnarray}
    \label{eq:brems_degN}
    \dot{q}_{\rm brems} = 3.4 \times 10^{33} \erg~\cm^{-3}~\s^{-1}
    \left(\frac{T}{10^{11}\K}\right)^8\left(\frac{\rho}{10^{13}
      g~\cm^{-3}}\right)^{1/3},
\end{eqnarray}
in the degeneracy regime of nucleons (Hannestad \& Raffelt 1998), and
\begin{eqnarray}
    \label{eq:brems_non-degN}
    \dot{q}_{\rm brems} = 1.5 \times 10^{33} \erg~\cm^{-3}~\s^{-1}
    \left(\frac{T}{10^{11}\K}\right)^{5.5}\left(\frac{\rho}{10^{13}
      g~\cm^{-3}}\right)^{2},
\end{eqnarray}
in the non-degeneracy regime of nucleons (Hannestad \& Raffelt 1998;
Burrows et al. 2000).

It is also known that the plasmon decay is effective in high densities
and high electron degeneracy region (Schinder et al. 1987). The decay
rate of the transverse plasmons which are normal photons interacting
with the electron gas through $\tilde{\gamma} \to \nu_e +
\anti{\nu}_e$ is estimated by Ruffert, Janka \& Sch\"afer (1996) as
\begin{eqnarray}
    \label{eq:plasmon}
    \dot{q}_{\rm plasmon} = 1.5 \times 10^{32}
    \erg~\cm^{-3}~\s^{-1}\left(\frac{T}{10^{11}\K}\right)^{9}
    \gamma_p^6 e^{-\gamma_p}(1 + \gamma_p) \left(2 +
      \frac{\gamma_p^2}{1 + \gamma_p}\right),
\end{eqnarray}
where $\gamma_p = 5.565 \times 10^{-2} \sqrt{(\pi^2 + 3 \eta_e^2)/3}$.
In particular, $\tilde{\gamma} \to \nu_e + \anti{\nu}_e$ is a dominant
process by a factor of $\sim 163$ compared with the other flavor ($\to
\nu_{\mu}\anti{\nu}_{\mu}$ or $\nu_{\tau}\anti{\nu}_{\tau}$ ). 

Note that in NPK only Eqs.~(\ref{eq:nucapture}) and
(\ref{eq:eeannihilation}) cases were considered in $\dot{q}_{\nu}$. To
see what component of the cooling rates mainly contributes to the
process, in Fig.~\ref{fig:cooling} we plot the most dominant one on
the ($\Sigma$, $T$) plane. It is interesting to note that the region
where advective energy transport dominates covers a rather wide, large
$\Sigma$ and high $T$ part of the $\Sigma-T$ diagram.  This is because
$Q_{\rm adv}$ ($\propto T^{16}$) has a stronger $T$ dependence than
$Q_{\rm rad}$ ($\propto T^4$) (shown later).  Also notice that
neutrino cooling is essential only at very high temperatures ($T
\gtrsim 10^{11} \K$) and very high matter density $\rho \gtrsim
10^{10}$ -- $10^{13} \g~\cm^{-3}$. The striped region represents the
place where heavy elements are produced which are expected from the
point of view in the equation of state of nuclear matter (Shen et al,
1998; Ishizuka, Ohnishi \& Sumiyoshi 2002). Here, we adopt the case of
the electron fraction $Y_e \simeq 0.1$ which the numerical simulations
of coalescing neutron stars predict, e.g., see Ruffert et al
(1997). Note that the left side of the boundary of the striped region
is just virtual one. Of course, the tail realistically continues to
the left further. As a result, there exist just a small amount of free
nucleons. Namely, the neutrino processes which we consider in this
study no longer become effective there.

Next, in Fig.~\ref{fig:tq_comb3}, we plot the heating and cooling
rates as a function of the temperature for some representative values
of surface density: (a) $\Sigma=10^{3.5}~ \g~\cm^{-2}$, (b)
$\Sigma=10^{8}~ \g~\cm^{-2}$, and (c) $\Sigma=10^{20}~
\g~\cm^{-2}$. In Fig.~\ref{fig:tq_comb3}(a), we find three
intersection points. They contain two points in which the energy
balance is realized due to the radiative or advective cooling
processes.  As for Fig.~\ref{fig:tq_comb3}(b), there exists only one
intersection point in which the energy balance is realized because of
the advective cooling processes. It is interesting that in
Fig.~\ref{fig:tq_comb3}(c), a new branch of the energy balance appears
at $T \gsim m_{\rm e} c^2 /k_{\rm B}$, since the neutrino-cooling
process becomes effective.  By plotting only the values at those
intersecting points on the ($\Sigma$, $T$) or ($\Sigma$, ${\dot M}$)
plane for various values of $\Sigma$, we obtain a sequence of the
thermal equilibrium solutions; that is the thermal equilibrium curve.

\subsection{Various time scales}
\label{sec:timescale}

Here it is necessary for us to check the time scale of the physical
processes of the neutrino cooling. The dynamical timescale is
represented by the accretion time,
\begin{eqnarray}
    \label{eq:tacc}
    \tau_{\rm acc} &\equiv& \frac1{\alpha}\sqrt{\frac{r^3}{G
    \mbh}}\left( \frac{r}{H}\right)^2, \\ 
    &\simeq& 1.3 \times 10^{-2} \ \s
    \left(\frac{\alpha}{0.1}\right)^{-1}\left(\frac{r}{4
      r_g}\right)^{3/2}\left(\frac{\mbh}{3 M_{\odot}}\right),
\end{eqnarray}
where we assumed that the disk half-thickness is approximately $H \sim
r/2$. On the other hand, the time scale of weak interaction of a
nucleon off the background electrons and positrons is estimated by
\begin{eqnarray}
    \label{eq:t_beta}
    \tau_{\beta} &\equiv& \left(\sigma_{Ne} n_e c \right)^{-1} \\
    &\sim& 1.0 \times 10^{-4} \ \s \left(\frac{T}{10^{11} \K}\right)^{-5},
\end{eqnarray}
where we assume that the cross section is $\sigma_{Ne} \sim G_{\rm
F}^2 (k_{\rm B}T)^2$, and the
electron number density is $n_e \sim (k_{\rm B}T/\hbar c)^3$.
Then, the condition $\tau_{\beta} \le \tau_{\rm acc}$ is realized when
the following relation is satisfied,
\begin{eqnarray}
    \label{eq:beta_equiv}
    T \ge 3.8 \times 10^{10} \K
    \left(\frac{\alpha}{0.1}\right)^{1/5}\left(\frac{r}{4
      r_g}\right)^{-3/10}\left(\frac{\mbh}{3 M_{\odot}}\right)^{-1/5}.
\end{eqnarray}
The interaction
rate of a neutrino off the background nucleons is roughly given by
\begin{eqnarray}
    \label{eq:gamma_nu}
    \Gamma_{N\nu} \sim G_{\rm F}^2 (\kb T)^2 n_Nc,
\end{eqnarray}
with the nucleon density $n_N$. Since $H \sim r/2$, the ratio of the
interaction time ($1/\Gamma_{N\nu}$) and the neutrino crossing time
($H/c$) is estimated by
\begin{eqnarray}
    \label{eq:mfp}
    \Gamma_{N\nu} \times H/c \simeq 42.6
    \left(\frac{T}{10^{11}\K}\right)^{2}
    \left(\frac{\rho}{10^{13}\g~\cm^{-3}}\right)
    \left(\frac{r}{4~r_g}\right)
    \left(\frac{\mbh}{3~M_{\odot}}\right).
\end{eqnarray}
Since the nucleon to electron-positron ratio is $n_N/n_e \sim
\order(10) (\rho/10^{13} \g~\cm^{-3})/(T/10^{11}\K)^{-3}$, we find
that the neutrino interaction is rapid enough to realize
$\beta$--equilibrium which we have assumed in the previous
subsections.
When $\Gamma_{N\nu} H/c$ is greater than unity, we should modify the
neutrino cooling rate in Eq.~(\ref{eq:neutrino_cooling}).  To
correctly estimate it in that case, we should divide the right-hand
side of Eq.~(\ref{eq:neutrino_cooling}) by $\Gamma_{N\nu} H/c$.
However, since the neutrino cooling rate becomes $\order(10^{2})$
times larger than the heating rate suddenly at intersection points
between the cooling and the heating rates, to estimate the equilibrium
curve, our simple treatments in the previous section are not so wrong
and the results are not changed.

Next, we estimate the vertical diffusion time of neutrinos. It is
given by
\begin{eqnarray}
    \label{eq:nu_diffuse}
    \tau_{\rm diff} &\equiv& \frac{(H/c)^2}{\Gamma_{N\nu}^{-1}}, \\ 
    &\simeq& 2.6 \times 10^{-3} \ \s
    \left(\frac{T}{10^{11}\K}\right)^{2}
    \left(\frac{\rho}{10^{13}\g~\cm^{-3}}\right)
    \left(\frac{r}{4~r_g}\right)^{2}
    \left(\frac{\mbh}{3~M_{\odot}}\right)^{2}.
\end{eqnarray}
Compared it with Eq.~(\ref{eq:tacc}), the condition of $\tau_{\rm diff}
\le \tau_{\rm acc}$ is translated into
\begin{eqnarray}
    \label{eq:cond_diff_short}
    T \le 2.3 \times 10^{11} \K
    \left(\frac{\alpha}{0.1}\right)^{-1/2} 
    \left(\frac{\rho}{10^{13}\g~\cm^{-3}}\right)^{-1/2} 
    \left(\frac{r}{4~r_g}\right)^{-1/4}  
    \left(\frac{\mbh}{3~M_{\odot}}\right)^{-1/2}.
\end{eqnarray}
Therefore, the neutrino cooling process is effective and realistic in
the equilibrium solutions of the accretion disk at a high temperature
$T \sim 10^{11} \K$ and a high density $\rho \sim 10^{13} \g~\cm^{-3}$
which will be studied in the next subsection.

\subsection{Thermal Equilibrium Solutions}
\label{sec:results}

According to the procedure discussed in
Sec.~\ref{sec:heating-cooling-rates}, we can find the thermal
equilibrium solutions; that is, intersecting points of the heating and
cooling curves. In Fig.~\ref{fig:sigma_t}
we plot them on the ($\Sigma$, $T$) plane for $r = 4~r_g$ (left panel)
and for $r = 40~r_g$ (right panel), respectively. In the parameter
region where neutrino cooling is effective, the dominant processes are
the electron capture rate ($N+e \to N' + \nu$).

\begin{deluxetable}{clllll}
\tablecolumns{8}
\tablewidth{0pc}
\tablecaption{$T$-$\Sigma$ and $\dot M$-$\Sigma$ Relations in Various Regimes}
\tablehead{
\colhead{} & \colhead{$p$} 
& \colhead{$Q^+$} 
& \colhead{$Q^-$} 
& \colhead{$T$-$\Sigma$ relation} & \colhead{${\dot
M}$-$\Sigma$~relation}}
\startdata
 I &$p_{\rm gas}$ &
$\alpha\Sigma Tr^{-\frac 32} \mbh^{\frac 12}$ &
$Q^-_{\rm rad} \propto \Sigma^{-1}T^4$ &
$\alpha^{\frac13}\Sigma^{\frac 23}r^{-\frac 12}\mbh^{\frac 16}$ &
$\alpha^{\frac43}\Sigma^{\frac 53}r\mbh^{-\frac 13}$
\\
II &$p_{\rm rad}$ &
$\alpha\Sigma^{-1}T^8r^{\frac 32}\mbh^{-\frac 12}$ &
$Q^-_{\rm rad} \propto \Sigma^{-1}T^4$ &
$\alpha^{-\frac 14}\Sigma^0r^{-\frac 38}\mbh^{\frac 18}$ &
$\alpha^{-1}\Sigma^{-1}r^{\frac 32}\mbh^{-\frac 12}$
\\
III &$p_{\rm rad}$ &
$\alpha \Sigma^{-1}T^8r^{\frac 32}\mbh^{-\frac{1}{2}}$ &
$Q^-_{\rm adv} \propto \alpha\Sigma^{-3}T^{16}r^{\frac{11}{2}}\mbh^{-\frac{5}{2}}$ &
$\alpha^0\Sigma^{\frac 14}r^{-\frac 12}\mbh^{\frac{1}{4}}$ &
$\alpha\Sigma r^{\frac 12}\mbh^{\frac{1}{2}}$
\\
IV  &$p_{\rm d,rel}$\tablenotemark{\dagger}\tablenotetext{\dagger}{
Here $p_{\rm gas}$ also contributes to the pressure secondly and
induces the $T$-dependence in $Q^-_{\rm adv}$.
}
&
$\alpha \Sigma^{\frac 97}r^{-\frac{27}{14}}\mbh^{\frac{9}{14}}$ &
$Q^-_{\rm adv} \propto \alpha \Sigma T^2r^{-\frac 12}\mbh^{-\frac{1}{2}}$&
$ \alpha^0\Sigma^{\frac 17}r^{-\frac{5}{7}}\mbh^{\frac{4}{7}}$ &
$\alpha\Sigma^{\frac 97}r^{\frac{15}{14}}\mbh^{-\frac{5}{14}}$
\\
V     &$p_{\rm d,rel}$ &
$\alpha \Sigma^{\frac 97}r^{-\frac{27}{14}} \mbh^{\frac{9}{14}}$ &
$Q^-_{\nu}\propto \eta_e^9T^9\Sigma^{\frac{1}{7}}r^{\frac{9}{7}}
\mbh^{-\frac{3}{7}}$ &
$\Sigma^{\frac{4}{7}}r^{-\frac{6}{7}}\mbh^{\frac{2}{7}}$ &
$\alpha \Sigma^{\frac 97}r^{\frac{15}{14}}\mbh^{-\frac{5}{14}}$
\\
\enddata
\end{deluxetable}

Using Eq.~(\ref{eq:Q_mdot}), we can also express
the mass accretion rate $\dot{M}$ at the intersection points as a
function of $\Sigma$.  The thermal equilibrium curves on the
 ($\Sigma$, $\dot M$) plane are plotted in Fig.~\ref{fig:sigma_mdot_3}.
Here, mass accretion rate is normalized by the critical mass accretion rate,
\begin{eqnarray}
    \label{eq:mdot_crit}
    \dot{M}_{\rm crit} = 16 L_{\rm Edd}/c^2,
\end{eqnarray}
where $L_{\rm Edd}$is the Eddington luminosity,
\begin{eqnarray}
    \label{eq:L_edd}
    L_{\rm Edd} = 4\pi GM_{\rm BH} m_{\rm p} c/\sigma,
\end{eqnarray}
with proton mass $m_{\rm p}$ and the cross section $\sigma$ of the
matter and the radiation through the electromagnetic scattering, i.e.,
$ L_{\rm Edd}/c^2 \simeq 7.3\times 10^{-17}M_{\odot}\cdot \s^{-1}
(\sigma/\sigma_{\rm T})^{-1} (M_{\rm BH}/M_{\odot})$ with the Thomson
cross section $\sigma_{\rm T} \simeq 0.6652 \ \mb$.

There are 5 distinct branches seen in these plots (see Table 1).  In
the lower-left parts of both figures, the equilibrium sequence has an
{\bf S} shape (see Abramowicz et al. 1988), which arises because of
changes in sources of pressure (gas and radiation pressure) and in
sources of cooling (radiation and advection).  In the upper branch of
the {\bf S} shape, advective energy transport takes over radiative
cooling.  Then, generated photons inside the disk take long time to go
out from the disk surface so that photons are advected inward and are
finally swallowed by a central black hole with accreting matter.
Then, disk luminosity is largely reduced.  Such trapping effects
should occur in branches III and IV with even higher $\dot M$.
Neutrino cooling is dominant only in the high temperature and density
regimes (branch V), in which neutrinos, instead of radiation, can
carry away the generated energy inside the disk. In branch V of Table
1, we wrote $Q^-_{\nu}$ in the case that only electrons are
degenerate. On the other hand, in the case that both electrons and
nucleons are degenerate, $Q^-_{\nu}$ becomes smaller by a factor of
$1/\eta_e^2$. Then, the thermal equilibrium solution is just realized
on the line of the condition of nucleon degeneracy. Namely, the
$T$--$\Sigma$ relation of the branch V coincides with that of nucleon
degeneracy.  In the intermediate branch (IV) still advective cooling
is dominant, however, note that entropy which is transported by
advection is contributed by non-degenerate nucleons, since entropy of
degenerate particles is little.

It is interesting to examine the stability of each branch. The
criterion for the thermal instability is
\begin{equation}
    \label{eq:thermal}
 \left(\frac{dQ^+}{dT}\right)_\Sigma > \left(\frac{dQ^-}{dT}\right)_\Sigma,
\end{equation}
whereas that for the viscous (secular) instability is
\begin{equation}
    \label{eq:viscous}
 \left(\frac{d{\dot M}}{d\Sigma}\right)_{Q^+=Q^-} < 0,
\end{equation}
(see, e.g., Kato, Fukue, \& Matsumoto 1998).
We understand immediately from Table 1 that
only branch II (radiation pressure-dominant standard disk branch)
is unstable for both modes
and all other branches, including the NDAF branch, are stable for 
any modes, as already noted by NPK.

\subsection{Properties of Neutrino-Cooled Disks}

>From the equilibrium solutions found in Sec.~\ref{sec:results}, we
derive the following expressions for the temperature and density on
the equatorial plane and for the corresponding mass-accretion rates as
functions of radius for a given value of $\Sigma$:
\begin{eqnarray}
    \label{eq:T_fit}
    T = 1.1 \times 10^{11}\K
    \left(\frac{\Sigma}{10^{20}\g~\cm^{-2}}\right)^{4/7}
    \left(\frac{r}{4~ r_g}\right)^{-6/7}
    \left(\frac{\mbh}{3~ M_{\odot}}\right)^{2/7},
\end{eqnarray}
\begin{eqnarray}
    \label{eq:rho_fit}
    \rho=2.3 \times 10^{13} \g~\cm^{-3}
    \left(\frac{\Sigma}{10^{20}\g~\cm^{-2}}\right)^{6/7}
    \left(\frac{r}{4~ r_g}\right)^{-9/7}
    \left(\frac{\mbh}{3~ M_{\odot}}\right)^{3/7},
\end{eqnarray}
and
\begin{eqnarray}
    \label{eq:mdot_fit}
    \dot{M} = 1.1 \times 10^{17} \dot{M}_{\rm crit}
             \left(\frac{\alpha}{0.1}\right)
            \left(\frac{\Sigma}{10^{20}\g~\cm^{-2}}\right)^{9/7}
            \left(\frac{r}{4~ r_g}\right)^{15/14}
            \left(\frac{\mbh}{3~ M_{\odot}}\right)^{-5/14}.
\end{eqnarray}
Note that NPK derived similar expressions but in terms of the disk
mass, $M_{\rm disk}$, and the size of the initial torus, $R_{\rm
out}$, which are roughly related to surface density as
\begin{equation}
  \Sigma \approx \frac{2M_{\rm disk}}{\pi b~R_{\rm out}^2}
         \sim \frac{10^{20}}{b} \g \ \cm^{-2}
             \left(\frac{M_{\rm disk}}{M_\odot}\right)
             \left(\frac{R_{\rm out}}{5\times10^6~{\rm cm}}\right)^{-2},
\end{equation}
where $b$ is a constant of the order of unity, which varies, depending
on the geometry of the initial torus and we assumed uniform surface
density distribution.  We thus understand that neutrino cooling
dominates only if a solar-mass material is contained within a disk
with a size of $\sim 5\times 10^6$ cm.  In other words, if the disk
mass is less or if the disk size is larger, neutrino cooling never
becomes dominant.  This constraint is more severe than that obtained
by NPK.

By comparing the left and right panels in Fig.~\ref{fig:sigma_t}, 
we notice that even at such a high surface density as 
$\Sigma \sim 10^{20} \g~\cm^{-2}$, 
the neutrino cooling does not become dominant at a
somewhat larger radius (i.e., $r \gtrsim 40~r_g$).
This also supports the claim by NPK that
NDAF solution only appears in a rather compact region around the center.

We can get a crude approximation for the strength of magnetic fields,
by using the above results.  The equi-partition field strength, for which
magnetic energy is equal to gas energy, is 
\begin{eqnarray}
\label{eq:B_eq}
B_{\rm eq} &=& \left(\frac{8\pi\rho k_B T}{m_{\rm p}}\right)^{1/2} \nonumber \\
          &\simeq& 7.3\times 10^{16} {\rm G} 
          \left(\frac{\alpha}{0.1}\right)^{1/12}
               \left(\frac{\Sigma}{10^{20}\g~\cm^{-2}}\right)^{5/7}
               \left(\frac{r}{4~r_g}\right)^{-15/14}
               \left(\frac{\mbh}{3~ M_{\odot}}\right)^{5/14}.
\end{eqnarray}
Therefore, the strength of the magnetic field is very likely to exceed
the critical value (e.g. M\'esz\'aros 1992),
\begin{equation}
   B_{\rm crit} \equiv \frac{m_{\rm e}^2c^3}{e \hbar}
         \simeq 4.4 \times 10^{13}~{\rm G},
\end{equation}
even if the fields strength is only 0.1 \% of the equi-partition value
(for which magnetic energy is only $10^{-4}$ \% of the particle
energy).  It is interesting to note that equilibrium $B$ value is
about the critical field for the proton,
\begin{equation}
   B_{\rm crit,p} = (m_{\rm p}/m_{\rm e})B_{\rm crit}
         \simeq 8 \times 10^{16}~{\rm G}.
\end{equation}
Then various QED (quantum-electrodynamics) effects should
manifest themselves (see the next section).

\section{Discussion}
\label{sec:discussion}

\subsection{CDAF and slim disk}
The present study is similar to that by NPK, although there are some
differences.  We newly take into account the effects of the electron
and nucleon degeneracy in neutrino cooling processes. The resultant
disk density and temperature differ in the neutrino-dominated regimes.
In the present study, we also take the slim-disk picture, instead of
the CDAF picture adopted by NPK, at moderately large $\Sigma$ (and
thus $\dot M$) ranges.  As mentioned above, the properties of the
hyper-critical accretion flow is not yet clear, since full
multi-dimensional simulations coupled with radiation hydrodynamics
(RHD) and possibly with magnetohydrodynamics (MHD) are finally
necessary to settle on this issue.  Probably convection plays a key
role there (Agol et al. 2001), but RHD simulations so far made in the
hyper-critical regimes are based on the approximation of the
flux-limited diffusion (e.g. Eggum, Coroniti, \& Katz 1988; Fujita \&
Okuda 1998), which may not hold within highly turbulent media.  Also,
magnetic reconnection, which is not easy to simulate in realistic
situations with extremely high magnetic Reynolds number, seems to play
an important role (see below).

Our conclusion basically agrees with that by NPK in that NDAF appears
at small radii only for high $M_{\rm disk}/r_{\rm out}^3$ cases.  NPK
concluded that models involving the mergers of BH-NS binaries and of
NS-NS binaries are favored to explain short GRBs because then the flow
is likely to be NDAF. However, our condition for the NDAF we found
is more stringent that they found and may not be generally satisfied.  

NPK also argue that the mergers of BH-WD binaries and BH-helium star
binaries are difficult to explain GRBs for the following reason.  Such
mergers will create disks with large dimensions, thus being unable to
reach the NDAF regimes but leading to the formation of CDAF.  Then,
convective motion mostly transports energy outward, thereby liberating
little energy at small radii.  It is thus unlikely to cause violent
explosions like GRBs.  However, we would like to point out that the
flow pattern is rather uncertain in the presence of magnetic fields.

If there are strong magnetic fields with poloidal (vertical)
components, for example, formation of outflow (or jet) is unavoidable
(e.g. Blandford \& Payne 1982; Kudoh, Matsumoto, \& Shibata 1998), and
such outflowing matter can produce enormous amount of synchrotron
photons. Magnetic reconnections, giving rise to intense flare
emission, may also occur.  That is, mergers of BH and WD or helium
star cannot be rejected as possible origins of GRBs.  More careful
discussion is needed regarding this issue.

\subsection{Proof of $p_{\rm rad} > p_{\rm gas}$ in  electron
non-degeneracy}
There are some discrepancies in the results between the present study
and NPK.  Those mainly come from a difference in the method. In NPK,
for example, they analytically studied the scaling law of physical
parameters in NDAF in two regimes, i.e., i) the gas pressure-dominated
regime, and ii) the degeneracy pressure-dominated regime.  Here, we
show that gas pressure never becomes dominant whenever electrons are
not degenerate at $T \gsim m_{\rm e} c^2/k_{\rm B}$.  Namely only when
electrons are degenerate, either gas pressure or degeneracy pressure
can become dominant, i.e., the case i) is not realized in electron
non-degeneracy.

Using Eqs.~(\ref{eq:rad_pressure})~and~(\ref{eq:gas_pressure}) with
the condition $n_{p} \sim n_{n}$, we obtain the relation,
\begin{eqnarray}
    \label{eq:pgas-prad}
    p_{\rm rad}/ p_{\rm gas} \simeq 13.02 \left( \frac{\rho}{10^{7}
      \g~\cm^{-3}}\right)^{-1}\left(\frac{T}{\mev}\right)^{3}.
\end{eqnarray}
On the other hand, the non-degeneracy condition for relativistic
electrons, see Eq.~(\ref{eq:deg_con_rela}), is represented by
\begin{eqnarray}
    \label{eq:deg_con_rela2}
     \left( \frac{\rho}{10^{7}
      \g~\cm^{-3}}\right)^{-1}\left(\frac{T}{\mev}\right)^{3} \gg 0.1142.
\end{eqnarray}
From Eqs.~(\ref{eq:pgas-prad})~and~(\ref{eq:deg_con_rela2}), we see
that the following relation is always realized,
\begin{eqnarray}
    \label{eq:prad_gg_pgas}
    p_{\rm rad} \gg p_{\rm gas},
\end{eqnarray}
in the non-degenerate electron regime at $T \gsim m_{\rm e} c^2/k_{\rm
B}$.

\subsection{Expected QED effects arising due to super-critical magnetic fields}
When the field strength exceeds the critical value ($\gtrsim
\order(10^{13})$~G), quantum electrodynamical processes become
important. Then, as it were, magnetic fields themselves behave as
{\lq\lq}particles" in the strongly magnetized vacuum, thus producing
several unique features. For example, strong magnetic fields induce
large energy splitting of Landau levels, large refractive indices of
photons, photon splitting effect, and so on (see, e.g., Adler 1971;
Shabad 1975; Melrose \& Stoneham 1976; Chistyakov et al. 1998; Kohri
\& Yamada 2002a).  In particular, the photon splitting effect is
interesting, when we consider the present model of the hyper-critical
accretion disk to a central engine of GRBs. That is because if the
rate of photon splitting ($\gamma \to 2\gamma$) is larger than that of
$e^\pm$ pair production ($\gamma + \gamma\to e^+ + e^-$), high energy
photons with their energy $E_{\gamma} \gg $MeV can lose kinetic energy
rapidly and produce a lot of soft photons ($E_{\gamma} \lsim 0.511
$MeV) without producing copious electron-positron pairs.  
Then, it may be possible to produce a fire ball as a source of GRB 
without requiring a large Lorentz factor.

Unfortunately, however, any concrete values of the photon splitting
rate near the threshold energy of the pair production has not been
calculated precisely except for some formal formulations (Adler 1971;
Shabad 1975; Melrose \& Stoneham 1976). The numerical study above the
threshold is currently underway (Kohri \& Yamada 2002b) and needs
further attention.

In addition, radiative processes in a ultra-magnetized plasma are
inherently complicated and we are still far away from having the basic
physics under control. At the enormous densities we are dealing with
the optical depth is huge, so radiative transfer must be solved.
Therefore, the above predictions may be premature at present. Only a
detailed radiative transfer calculation, which is outside the present
capabilities of numerical codes, could address this point.

\subsection{Detectability of neutrinos from NDAFs}
As we have shown in the previous sections, we expect that a lot of
neutrinos would be emitted from NDAF.  It is thus of great importance
to check whether the signals of the neutrinos are detectable or not in
GRB events. Moreover, it is also informative to find some correlations
between neutrino emission and photon emission from GRBs in terms of
the observed times and/or directions to the source of the neutrinos.

Such investigation has already been initiated by Nagataki and Kohri
(2001), who computed the neutrino luminosity and its detectability,
properly considering the time evolution of the central BH. Their
calculations are based not on the merger models but on the collapsar
model, and they adopt the simple analytical fitting formula to the
hyper-critical (slim) disk model taken from Fujimoto et al. (2001).
(Basic numbers are in good agreement with those obtained by the
numerical evaluations by Popham et al 1999.)  Their conclusion is that
we will be able to marginally detect the neutrinos from collapsars
occurring with ``TITAND,'' a next-generation multi-megaton water
Cherenkov detector (Suzuki 2001).  The first phase of TITAND is planed
to be with 2 Mt water inside. In addition, we will be also able to
know the direction to the source of neutrinos in TITAND.  For the
total number of emitted $\anti{\nu}_e$, we roughly evaluate
\begin{eqnarray}
    \label{eq:N_nu}
    N_{\anti{\nu}_e} &\sim& \frac{\dot{q}_{Ne} V \Delta t}
    {\anti{E}_{\anti{\nu}_e}} \\ 
    &\sim& 10^{57} \left(\frac{\dot{q}_{Ne}}{10^{35}
      \erg~\cm^{-3}~\s^{-1}}\right)
    \left(\frac{\anti{E}_{\anti{\nu}_e}}{10 \mev}\right)^{-1}
    \left(\frac{V}{10^{20} \cm^{3}}\right) \left(\frac{\Delta
      t}{10^{-3} \s}\right),
\end{eqnarray}
with their mean energy $\anti{E}_{\anti{\nu}_e}$ to be emitted during
one GRB event, where $\dot{q}_{Ne}$ is estimated at $T \simeq 10
\mev/\kb$ and $\rho = 10^{13} \g~\cm^{-3}$, $V$ is the volume of the
emitting region, and $\Delta t$ is the duration. For the event number
of $\anti{\nu}_e$'s,
\begin{eqnarray}
    \label{eq:event_number}
    R_{\rm event} \sim 10 \left(\frac{d}{3 \ {\rm Mpc}}\right)^{-2}
    \left(\frac{N_{\anti{\nu}_e}}{10^{57}}\right)
    \left(\frac{\anti{E}_{\anti{\nu}_e}}{10 \ \mev}\right)^{2}
    \left(\frac{V_{\rm H_2O}}{2 \ {\rm Mt}}\right)
\end{eqnarray}
will be detected by TITAND, where $d$ is the distance from the source
and $V_{\rm H_2O}$ is the volume of water.  Hence, we optimistically
expect to detect neutrinos from NDAFs, if GRB occurs within a distance
of 3 Mpc.  If detected, neutrino signals and correlations with GRB
will be able to make clear the unknown explosion mechanism of GRBs.
In order to derive the reliable estimations on neutrino flux, however,
we should perform detailed computations as was done by Nagataki and
Kohri (2001).  This is left as future work.

\section{Conclusion}
\label{sec:conclusion}

We study the properties of hyper-critical accretion flow and for
surface density exceeding about $10^{20}$ g cm$^{-2}$, which realizes
when about a solar mass material is contained within a disk with a
size of $\sim 5\times 10^6$ cm, we find the following unique features:
\begin{enumerate}

\item Radiation luminosity of such flow is practically zero due to
    significant photon trapping, although mass accretion rate
    enormously exceeds the critical rate, ${\dot M}\gg L_{\rm
    Edd}/c^2$.
\item Neutrino-cooling dominates over advective cooling. Thus the
    flow can cool via neutrino emission.  We expect to detect
    neutrinos from GRBs with the next-generation, multi-megaton water
    Cherenkov detector in future.
\item Electron degeneracy pressure dominates over gas and radiation
    pressure, and the degenerate electron definitely influences the
    processes of neutrino emission. This feature is distinct from that
    of the solution found by NPK.  
\item The disks are stable both against thermal and viscous
    instabilities, even when we take into account electron degeneracy.
\item Magnetic field strength exceeds the critical value, even if we
    take only 0.1\% of the equi-partition value.  Then, photon splitting
    may occur, producing significant number of photons of energy below
    $\sim 511$ keV, thereby possibly suppressing e$^\pm$ pair
    creation.  However, quantitative discussion is left as future
    work.
\end{enumerate}

\acknowledgements
We are grateful to R. Narayan, Shoichi Yamada and S. Nagataki for
useful discussions and comments.  The comments by the anonymous
referee are also suggestive and helpful in making the final draft.  We
also thank the Yukawa Institute for Theoretical Physics at Kyoto
University for the YITP workshops YITP-W-01-13? on ``Gamma-Ray
Bursts,'' and YITP-W-01-17 on ``Black Holes, Gravitational Lens, and
Gamma-Ray Bursts,'' in which this work was initiated and completed.
This work was supported in part by the Grants-in Aid of the Ministry
of Education, Science, Sports, and Culture of Japan (13640238, SM).
Numerical computation in this work was carried out at the Yukawa
Institute Computer Facility.



\newpage

\centerline{{\vbox{\epsfxsize=13.0cm\epsfbox{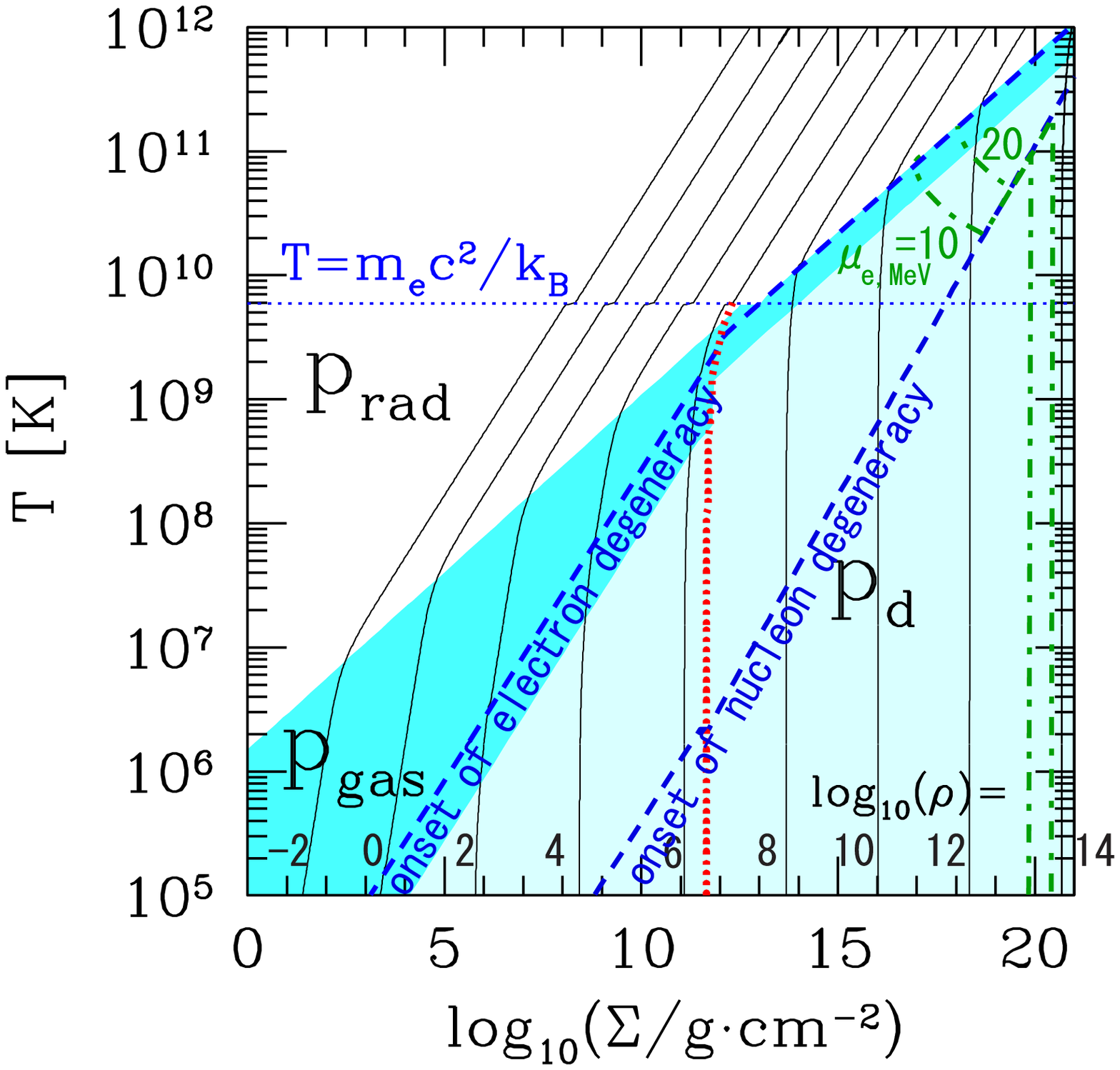}}}}
\figcaption{
\label{fig:rho_cont}
Contours of the matter density on the ($\Sigma$, $T$) plane for case
that the black-hole mass is $M_{\rm BH} = 3~ M_{\odot}$ and $r
=4~r_{\rm g}$, where $r_{\rm g}$ is the Schwarzschild radius.  The
solid lines denote the contours of the matter density for
$\log(\rho/\g~\cm^{-3})$ = --2, 0, 2, $\cdots$, and 14.  The dark (or
light) shadowed region represents that $p_{\rm gas}$ ($p_{\rm d}$)
dominantly contributes to the total pressure, while the white region
represents that $p_{\rm rad}$ is dominant. The left and right thick
dashed lines, respectively, denote the loci where electron and nucleon
degeneracy begin to take place.  Namely, electrons (or nucleons) are
degenerate because of higher density on the right side of the left
(right) dashed line.  The horizontal dotted line denotes the
temperature which corresponds to the electron mass $m_{\rm e}$. The
thick dotted line denotes the boundary where electrons become
relativistic due to higher energy density ($\rho \gtrsim 2 \times 10^6
\g~\cm^{-3}$) while $T \lesssim m_e c^2/k_{\rm B}$. The dot-dashed
lines represent the chemical potential of electron in units of MeV,
$\mu_{e,{\rm MeV}}$ = 10, and 20, from the lower left to the upper
right.
}
\vspace{0.5cm}

\newpage
\vspace{0.5cm}
\centerline{{\vbox{\epsfxsize=13.0cm\epsfbox{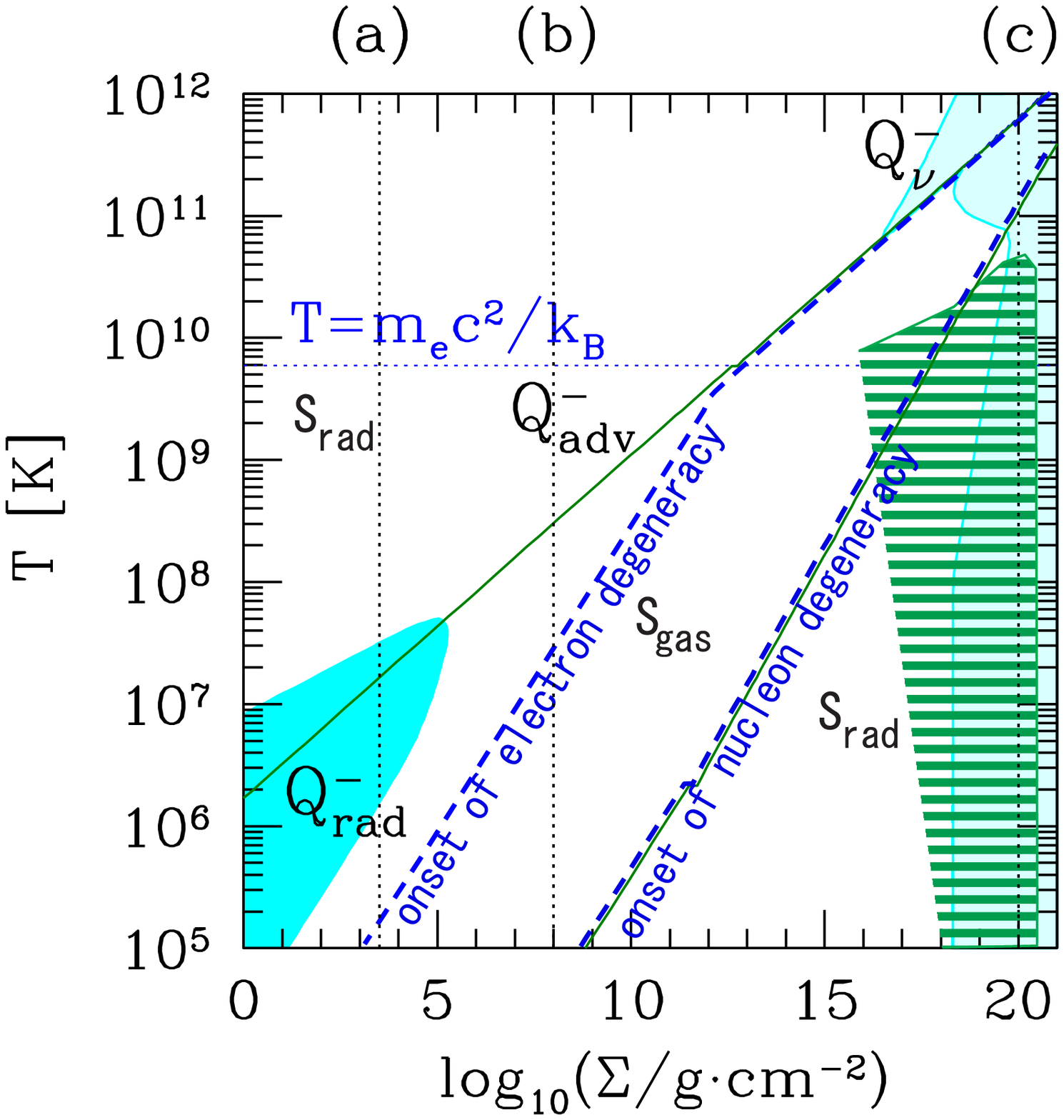}}}}
\figcaption{
\label{fig:cooling}
Plot of the most dominant component in the cooling rates on the
($\Sigma$, $T$) plane. The adopted parameters are the viscosity
parameter, $\alpha = 0.1$, the black-hole mass, $M_{\rm BH} = 3~
M_{\odot}$, and $r = 4~r_{\rm g}$.  The dark (light) shadowed region
and white region, respectively, represent the places where the
radiative (neutrino) cooling and the advective cooling is the dominant
process. On the other hand, the region between the two solid lines
represents the place where the gas entropy $s_{\rm gas}$ mainly
contributes to the total entropy. The striped region represents the
place where heavy elements are produced, and there exist just a small
amount of free nucleons. The vertical dotted lines, (a), (b), and (c)
denote the representative values of $\Sigma$, at which heating and
cooling rates are plotted in Fig.~\ref{fig:tq_comb3}. The dashed and
horizontal dotted lines are the same as those in
Fig.~\ref{fig:cooling}.
}

\newpage
\vspace{0.5cm}
\centerline{{\vbox{\epsfxsize=17.cm\epsfbox{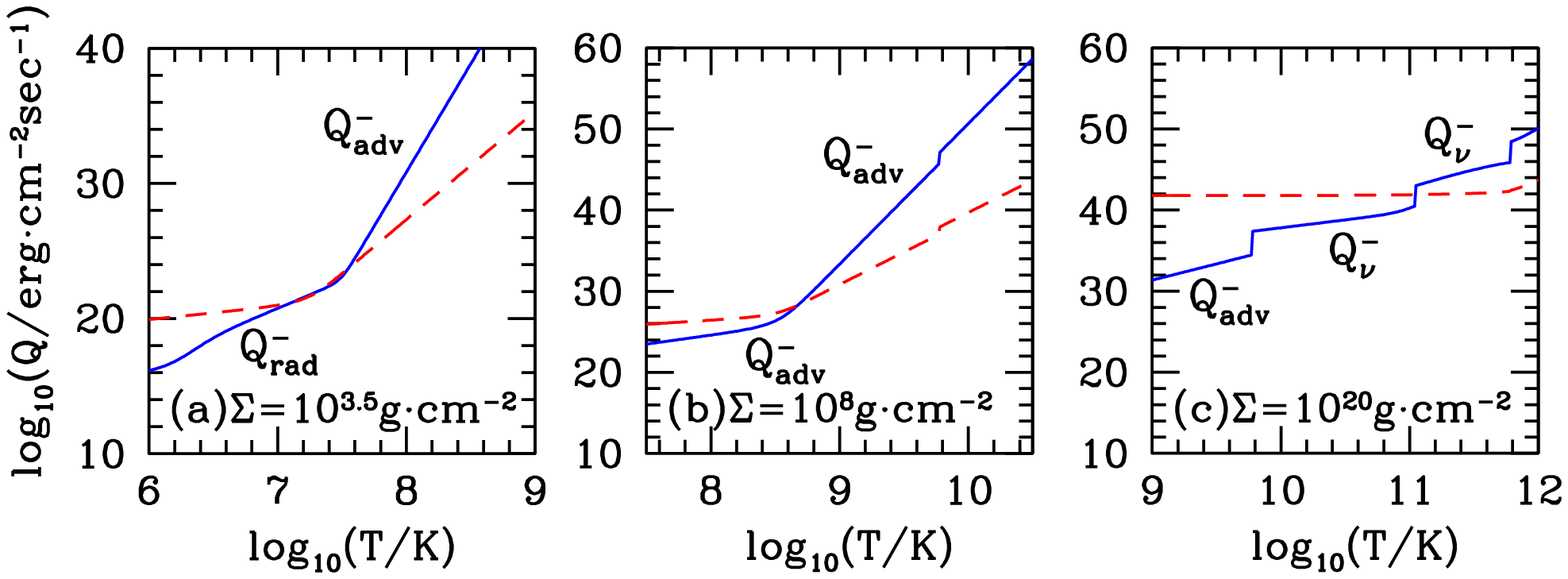}}}}
\figcaption{
\label{fig:tq_comb3}
Heating and cooling rates as functions of temperature
for a fixed surface density; from left to right,
(a) $\Sigma=10^{3.5} ~\g~\cm^{-2}$,
(b) $\Sigma=10^{8} ~\g~\cm^{-2}$, and 
(c) $\Sigma=10^{20}~\g~\cm^{-2}$, respectively. 
The other parameters are the same as those in Fig.~\ref{fig:cooling}.
}

\vspace{0.5cm}

\centerline{{\vbox{\epsfxsize=15.cm\epsfbox{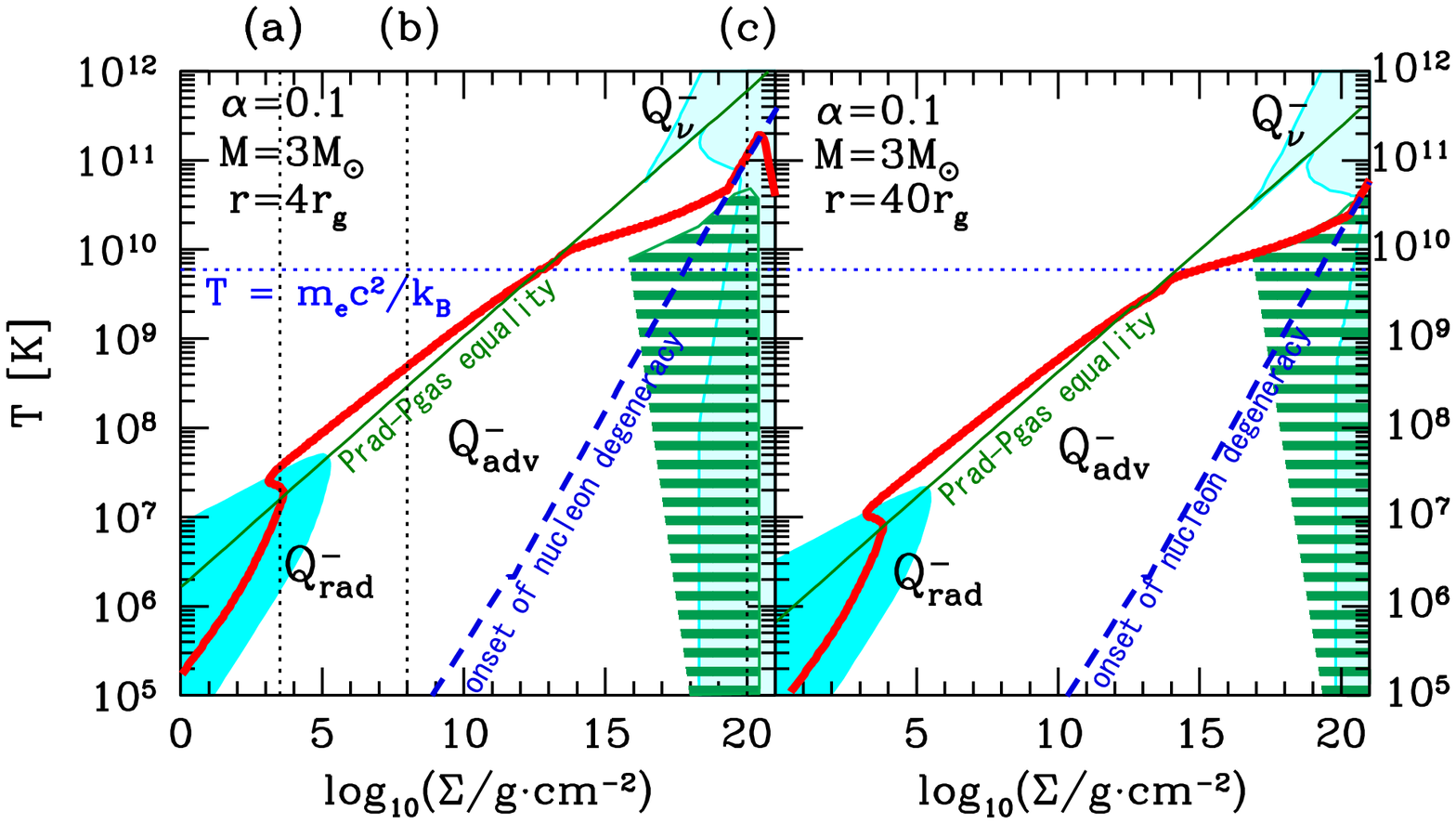}}}}
\figcaption{
\label{fig:sigma_t}
Thermal equilibrium curves (by the thick solid line) on the ($\Sigma$,
$T$) planes at radial distances of $r = 4~r_{\rm g}$ (left panel) and
$r = 40~r_{\rm g}$ (right panel).  The adopted parameters are $\alpha
= 0.1$ and $M_{\rm BH} = 3~M_{\odot}$.  The vertical dotted lines (a),
(b), and (c) denote the three representative values of $\Sigma$
adopted in plotting Fig.~\ref{fig:tq_comb3}.  The thin line indicates
the loci where gas pressure equals radiation pressure.  The dashed
line indicates the loci where nucleon degeneracy sets out. We also
indicate dominant source of cooling by shaded zone as in
Fig.~\ref{fig:cooling}. In the striped region, there exist just a
small amount of free nucleons, and the neutrino cooling does not work
well.
}
\vspace{0.5cm}

\newpage

\vspace{0.5cm}
\centerline{{\vbox{\epsfxsize=13.cm\epsfbox{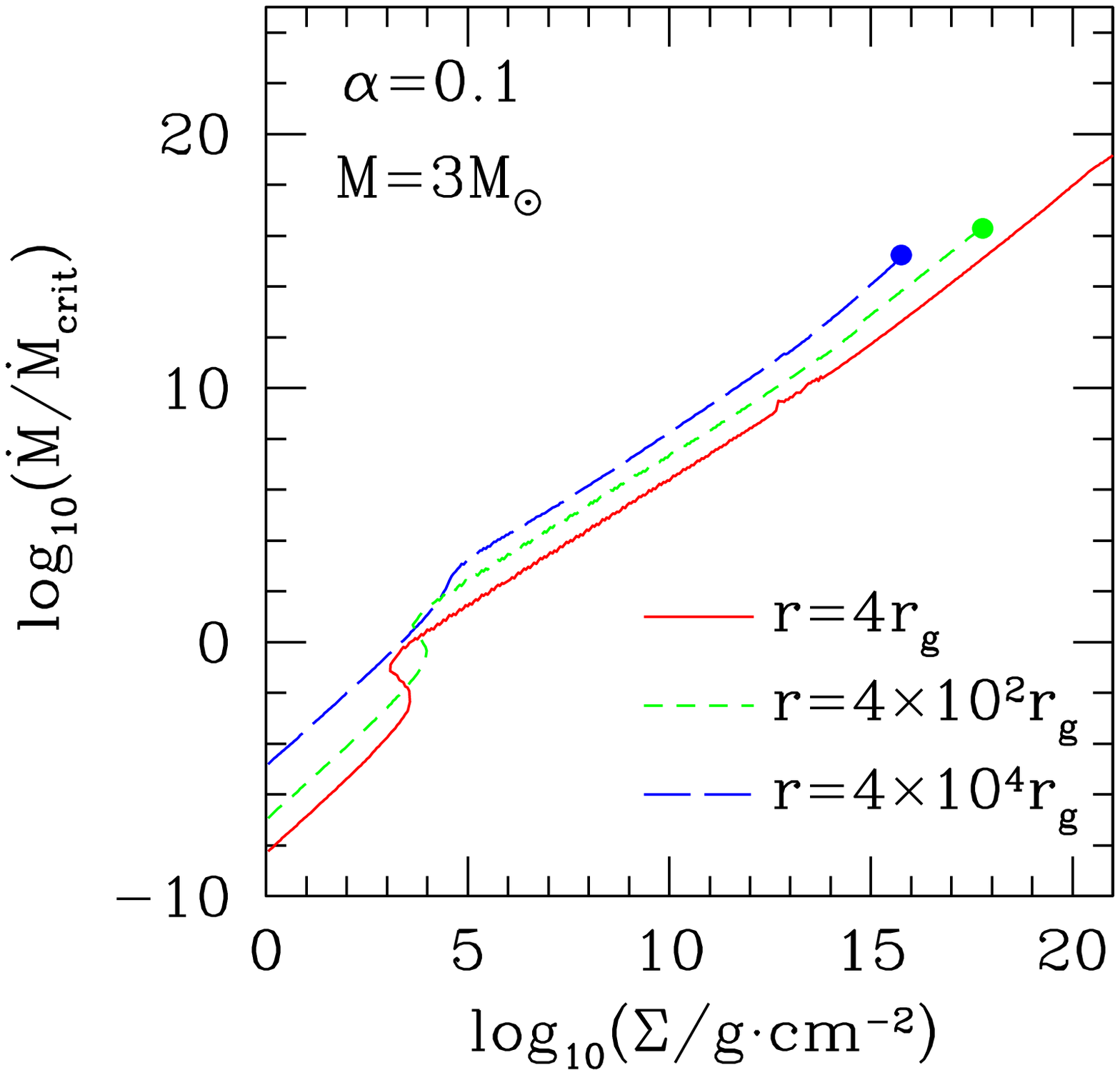}}}}
\figcaption{
\label{fig:sigma_mdot_3}
Thermal equilibrium curves on the ($\Sigma$, ${\dot M}$) plane
at $r = 4~r_{\rm g}$ (solid line) , $400~r_{\rm g}$ (dashed line),
and $4\times10^4~r_{\rm g}$ (long-dashed line).
All other parameters are the same as
those in Fig.~\ref{fig:sigma_t}.  
The solid circles at the upper ends indicate the places 
of $H=r$, above which no physical solution exists.
}
\vspace{0.5cm}

\end{document}